\documentclass[twocolumn,showpacs,preprintnumbers,amsmath,amssymb,prb]{revtex4}

\usepackage{graphicx}
\usepackage{dcolumn}
\usepackage{bm}
\begin{document}


\title{Taming the Rugged Landscape: Techniques for the Production,
Reordering, and Stabilization of Selected Cluster Inherent Structures }

\author{Dubravko Sabo}
\author{J. D. Doll}
\affiliation{Department of Chemistry, \\ Brown University, \\ Providence,
RI 02912, USA}

\author{David L. Freeman}
\affiliation{Department of Chemistry, \\ University of Rhode Island,
\\ Kingston, RI 02881, USA}

\date{\today}

\begin{abstract}
We report our studies of the potential energy surface (PES) of selected binary
Lennard-Jones clusters. The effect of adding selected impurity atoms 
to a homogeneous cluster is explored. Inherent structures and transition
states are found by combination of conjugate gradient and 
eigenvector-following methods while the topography of the PES is mapped 
with the help of a disconnectivity analysis. We show that we can controllably
induce new structures as well as reorder and stabilize existing structures
that are characteristic of higher-lying minima.
\end{abstract}

\pacs{82.20.Wt,02.60.Pn}

\maketitle

\section{Introduction} \label{sec:intro}

The minimization/optimization problem is one of the more ubiquitous and
challenging in computational science \cite{PRESTEU}.
Central to researchers in the physical
sciences and engineering, this problem is also of primary importance to
social, biological, and economics investigators.

Driven in large measure by such widespread interest, there has been appreciable
progress on the minimization problem. Especially notable have been algorithmic
advances in the form of annealing and stochastic relaxation approaches 
\cite{PRESTEU,KIRGEVE,SCHERAGA86,SCHERAGA87,FAKEN99,BERNE00}
as well as basin-hopping techniques \cite{WALES97B,WALES99S}.
In both classical and quantum form, these methods
offer valuable, complementary alternatives to traditional, gradient or 
pseudo-gradient approaches \cite{PRESTEU}.

In addition to algorithmic developments relevant to the minimization problem,
there have also been notable advances in the tools to classify and analyze
the topography of the underlying objective functions. In chemical applications,
the principal focus of the remainder of our discussion, the objective
function of interest is typically a specified potential or free energy surface.
Following Stillinger and Weber \cite{STILLINGER83,STILLINGER84}, 
it is useful to perform
an ``inherent structure'' decomposition of the associated configuration space
by employing the minima (local and global) of this surface. These inherent
structures, their relative orderings, and their connectivity provide
important information concerning the structure, function, and dynamics of
the associated physical system. Disconnectivity analysis introduced by
Czerminski and Elber \cite{ELBER90}, discussed by
Becker and Karplus \cite{KARPLUS97} and developed by, 
among others, Wales, Doye and Miller \cite{WALES99A,WALES99B,WALES99C}
has proved especially valuable with respect to these latter tasks.

As evidenced by the development of classical and quantum annealing methods,
there is an important interplay between minimization and the Monte Carlo
sampling problem. Both applications, for example, are concerned with
overcoming barriers that inhibit the interconversion or isomerization 
of the various inherent structures of the problem. Consequently, 
developments in one field contain implications for developments in the
other. Advances in rare event sampling methods, such as 
J-walking \cite{FREEMAN92,JORD93C} and
parallel tempering \cite{PARISI92,THOMPSON95,FREEMAN00A,FREEMAN00B}
methods, thus contain implications for the minimization problem.

The field of atomic and molecular clusters has been and continues to be
an important test bed for the development and application of minimization
and analysis methods. Utilizing the methods outlined above,
researchers have produced a relatively coherent picture of the relationship
between the nature of underlying PES and the physical properties of the
associated systems. For example, from the studies by Berry {\it et al.}
\cite{BERRY86,BERRY90,BERRY93,BERRY94},
the single component studies of Wales {\it et al.}
\cite{WALES99A,WALES99B,WALES99C},
the mixed LJ cluster studies of Jordan {\it et al.} \cite{JORD02}
as well as the research of others \cite{ABRAHAM78,FRANTZ96,FRANTZ97,
SERRA97}, 
we have begun to understand the nature of systems for which the lowest
inherent structures can or cannot be readily located. 

In present paper we would like to build upon advances in the minimization
problem by effectively turning the logic ``upside down''. That is,
instead of asking what we have to do in order to locate or sample
the global minimum of a specified potential energy surface, we wish
instead to ask how we might go about controllably inducing new structures
as well as reordering and stabilizing existing structures that are 
characteristic of higher-lying local minima. Basically, we seek to utilize
what we have learned about what it takes to {\em avoid} local minima 
to instead {\em controllably produce} them.

In principle, one can envision efforts involving both thermodynamic and
kinetic approaches. In the present work we shall focus principally on 
the thermodynamic issues. Furthermore, we shall limit the discussion
in the present work to applications involving clusters. As discussed
elsewhere \cite{JELL99}, 
clusters are of appreciable technological importance,
are valuable as prototypes for the study of the properties of extended
systems, illuminate issues related to the size-dependence of selected
physical properties, and provide valuable test beds for the development
and application of emerging computational techniques. This combination
of formal, computational and technological interest has produced a
vast and growing cluster literature \cite{FREEMAN96,JELL99,KNICKEL99}.

The remainder of the paper is organized as follows: 
In Section~\ref{sec:comput} we outline the computational details
of the present study.  We discuss the methods we use to determine
the inherent structures and transition states of a specified
cluster's potential energy surface. Using these methods, we examine 
specific results for two prototype systems in Section~\ref{sec:numres}. 
These particular results are
designed to demonstrate ``proof of principle'' with respect to
the basic objectives of the present study for selected systems.  
Finally, in Section~\ref{sec:conclude} we summarize our results 
and speculate about likely future research
directions.

\section{Computational Details} \label{sec:comput}

The present Section describes the computational details of our
investigations involving binary clusters of the form $\mathrm{X_nY_m}$. Our
overall interest will be to explore the extent to which we can
utilize the ``adatoms'' (i.e. the Y-system) to induce, reorder and
stabilize selected inherent structures in the ``core'' X-system. While
one can easily imagine applications involving both more and more
complex components, we feel these relatively simple, two-component
clusters are a convenient starting point for an initial study of the
issues we raise.

We shall assume in what follows that the total potential energy is
composed of a pairwise sum of Lennard-Jones interactions. Specifically,
we assume that the total potential energy, $\mathrm V_{tot}$, for an 
N-particle system is given by

\begin{equation}
\label{2.1}
V_{tot} = \sum_{i<j}^{N} v_{ij}(r_{ij}),
\end{equation}
where the pair interaction as a function of the distance between
particles i and j, $r_{ij}$, is given by

\begin{equation}
\label{2.2}
v_{ij}(r_{ij}) = 4\epsilon_{ij}\  
      [( \frac{\sigma_{ij}}{r_{ij}})^{12}
     -( \frac{\sigma_{ij}}{r_{ij}})^{6}] .
\end{equation}
In Eq. (\ref{2.2}) the constants $\epsilon_{ij}$ and $\sigma_{ij}$ are
the energy and length-scale parameters for the interaction of
particles i and j.

For a two-component system, we must specify both the ``like'' 
(X-X, Y-Y) as well as the ``mixed'' (X-Y) interactions. With an eye
toward studying trends in the results as opposed to results for 
particular physical systems, it is convenient to reduce the number
of free parameters. To do so, we shall assume in the present study
that the ``mixed'' Lennard-Jones values are obtained from the ``like''
Lennard-Jones parameters via usual combination rules \cite{COMBINE}

\begin{equation}
\label{2.3}
\sigma_{_{XY}}= \frac{1}{2}(\sigma_{_{XX}}+\sigma_{_{YY}})
\end{equation}

\begin{equation}
\label{2.4}
\epsilon_{_{XY}}= \sqrt{\epsilon_{_{XX}}\epsilon_{_{YY}}} .
\end{equation}
Furthermore, we note that with the mixed Lennard-Jones parameters
specified as in Eqs.(\ref{2.3}) and (\ref{2.4}), the inherent structure
topography of the ``reduced'' potential energy surface of the binary
system (i.e. $V_{tot}/\epsilon_{_{XX}}$) is a function of only two
parameters, ($\sigma, \epsilon$), the ratios of the corresponding
adatom/core length and energy parameters

\begin{equation}
\label{2.5}
\sigma= \sigma_{_{YY}}/\sigma_{_{XX}}
\end{equation}

\begin{equation}
\label{2.6}
\epsilon= \epsilon_{_{YY}}/\epsilon_{_{XX}} .
\end{equation}

If necessary for a discussion of a specific physical system, the
absolute bond lengths, energies, activation energies, etc. can be
obtained from the corresponding ``reduced'' results by a simple
rescaling with the appropriate core-system Lennard-Jones parameters.

The computational task in our study is thus one of exploring and
characterizing the (reduced) potential energy surface of our
binary cluster systems as a function of the number of (core, adatom)
particles, (n,m), and for given ($\sigma,\epsilon$) ratios. In
typical applications the lowest $\mathrm N_{IS}$ inherent structures and the
associated disconnectivity graphs are determined. For the applications
reported here, $\mathrm N_{IS}$ is generally of the order of a few hundred 
(thousand) or less. Depending on the size of the cluster, inherent structures
are found either via conjugate gradient methods starting from randomly
chosen initial configurations, or by more systematic surface exploration
methods such as those outlined by Wales and co-workers \cite{WALES99B}
and by Jordan {\it et al.} \cite{JORD93B}.
In all cases, the inherent structures that are located are 
confirmed to be stable minima via a standard Hessian analysis. To
reduce the chance we miss particular local or global minima, we monitor
the number of times individual inherent structures are found and
demand that each of the N$_{IS}$ inherent structures be located a
minimum number of times (at least 10) before we terminate our search.
Once we are satisfied we have located the relevant inherent structures,
transitions states linking these stable minima are obtained using the
eigenvector following methods outlined by Miller and Cerjan \cite{MILLER81}
and further developed by Simons {\it et al.} \cite{SIMONS83,SIMONS85,SIMONS90},
Jordan {\it et al.} \cite{JORD93B} and Wales \cite{WALES94}.
Finally, with the requisite inherent structures and
barriers in hand, we perform a disconnectivity analysis using methods
outlined by Czerminski and Elber \cite{ELBER90}, 
Becker and Karplus \cite{KARPLUS97} and Wales {\it et al.}
\cite{WALES99A} .

\section{Numerical Results} \label{sec:numres}

In the present Section, we wish to illustrate the general themes we
introduced in Section~\ref{sec:intro} . 
We do so by demonstrating that we can accomplish
three basic objectives. Specifically, we show that by adding selected
``impurity'' atoms to bare ``core'' systems, we can:

1. induce new ``core structures'' 

2. reorder the energies of existing core inherent structures, and 

3. stabilize selected inherent structures by controlling the activation 
energies that determine their isomerization kinetics.

\begin{figure}[!tbp]
\includegraphics[clip=true,width=3.2cm]{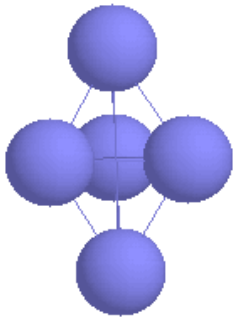}
\caption{\label{fig:core5} The only stable inherent structure for X$_5$ 
LJ cluster. Its energy (in units of the LJ well depth) is -9.104.}
\end{figure}

\begin{figure}[!htbp] \centering
  \begin{tabular}{@{}cc@{}}
    \includegraphics[width=4.2cm,clip=true]{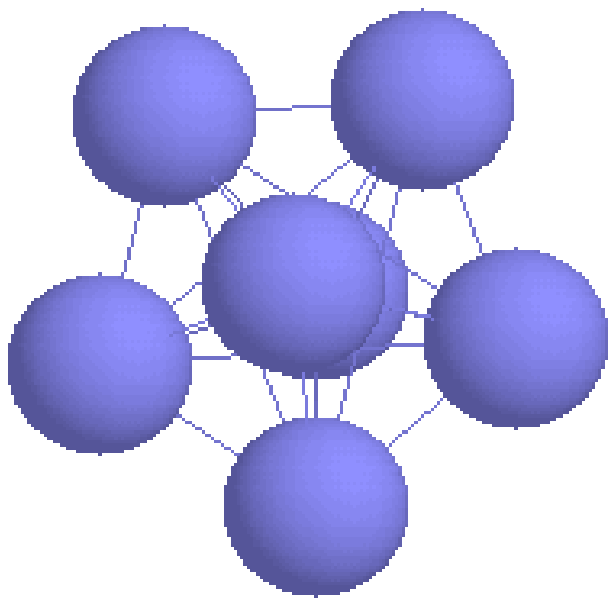} & 
    \includegraphics[width=4.2cm,clip=true]{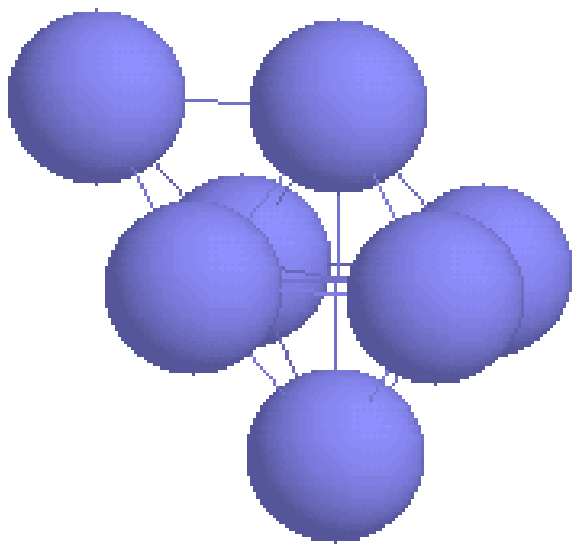} \\
    (a) & (b) \\
    \includegraphics[width=4.2cm,clip=true]{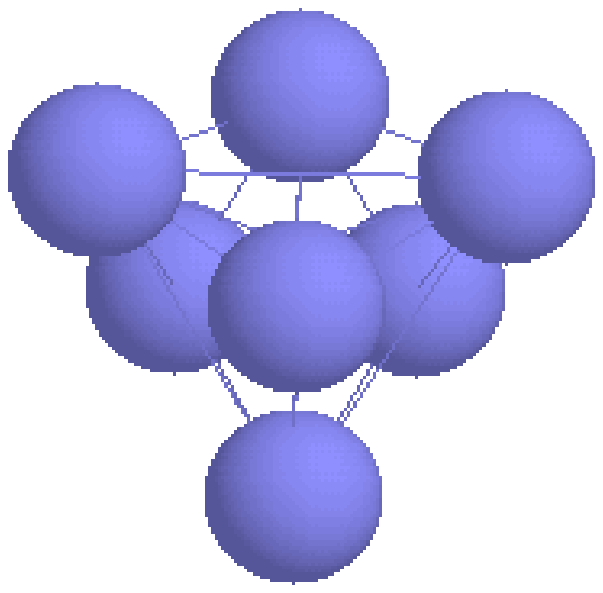} & 
    \includegraphics[width=4.2cm,clip=true]{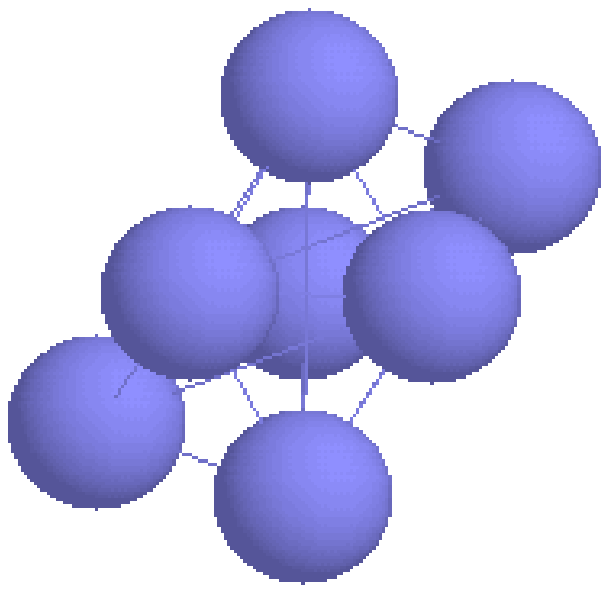} \\
    (c) & (d) \\
        \end{tabular}
\caption{\label{fig:core7}
The four, energetically distinct, stable inherent structures for
X$_7$ LJ cluster. The energies (in units of the LJ well depth) are: (a)
-16.505, (b) -15.935, (c) -15.593, (d) -15.533.}
\end{figure}

For purposes of illustration, we shall examine numerical results for a few,
simple Lennard-Jones systems involving five and seven core atoms, systems
well-known from previous studies to have one and four energetically distinct
inherent structures, respectively. The inherent structures and their
associated energies for these core systems are illustrated in Fig.
~\ref{fig:core5} and ~\ref{fig:core7}.

\begin{figure}
\includegraphics[clip=true,width=8.5cm]{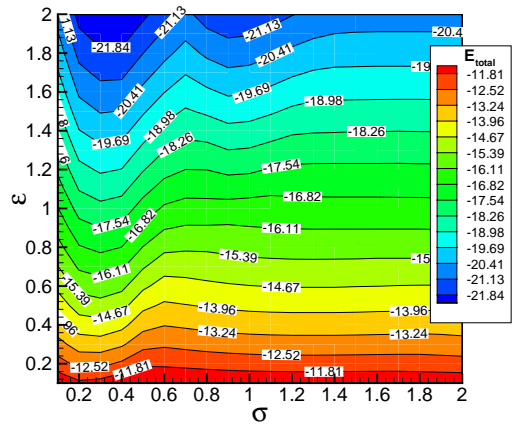}
\caption{\label{fig:totPES52}
$\mathrm E_{tot}(\sigma,\epsilon)$ (c.f.Eqs.\ref{2.1}, \ref{2.5} and \ref{2.6})
for the $\mathrm{X_5Y_2}$ system. Note the relative lack of structure in the
($\sigma$,$\epsilon$) variation of the total cluster energy.}
\end{figure}

We first consider mixed clusters of the generic type X$_5$Y$_2$. Here two
impurity Y-atoms are added to the parent, five-atom X-core. We have chosen
this system because it builds upon the very simple five-atom core, a system
that has only a single inherent structure, and because the total system has
a total of seven atoms, a magic number for icosahedral growth in homogeneous
systems. Using the techniques of Section II, we then determine the lowest
several inherent structures for a range of ($\sigma$,$\epsilon$) [c.f.
Eq.(\ref{2.5}) and Eq.(\ref{2.6})]. 
As can be seen from Fig.~\ref{fig:totPES52}, the total
potential energy [Eq.(\ref{2.1})] of the lowest inherent structure for the
X$_5$Y$_2$ system shows no appreciable structure as a function of the
($\sigma$,$\epsilon$) parameters. 

\begin{figure}[!htbp]\centering
\includegraphics[clip=true,width=8.5cm]{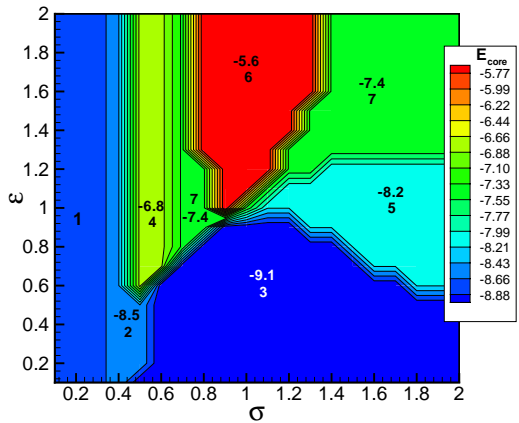}
\caption{\label{fig:corePES52}
$\mathrm E_{core}(\sigma,\epsilon)$ for the $\mathrm{X_5Y_2}$ system.
Here the ``core'' energy is defined as that portion of the potential energy
arising from only the core-core atom interactions. Unlike the total energy,
the $\mathrm(\sigma,\epsilon)$ variation of the core cluster energy exhibits
relatively well-defined regions. The labels of each of these regions in
the figure correspond to the distinct core structures shown in
Fig.~\ref{fig:coreE52}.}
\end{figure}

On the other hand, we see in Fig.~\ref{fig:corePES52}
that the core potential energy, defined as the
potential energy of interaction for only the core X-atoms, of the minimum
(total) energy cluster clearly breaks into extended regions, each 
corresponding to a well-defined core structure. The reader should notice
that each region in Fig.~\ref{fig:corePES52} contains the same ``kind''
of core structure but their core energies are slightly different. We have
chosen a single ``average'' core energy value to represent all energies in the 
corresponding domain for plotting convenience.

The distinct core structures, shown in Fig.~\ref{fig:corePES52}, have
been identified by examining their core energies (E$_{core}$) and their
principal moments of inertia. For each structure a triplet of
values (E$_{core}$, I$_2$, I$_3$) has been associated, where I$_2$ and
I$_3$ are the moments of inertia about the principal axes 2 and 3, 
respectively. We have defined I$_2$ and I$_3$ in the following way:
I$_2$=I$^{'}_2$/I$^{'}_1$, I$_3$=I$^{'}_3$/I$^{'}_1$ where I$^{'}_1$,
I$^{'}_2$ and I$^{'}_3$ are the principal moments of inertia obtained
by diagonalizing the inertia tensor of the system. If the triplet of 
values has not been sufficient to identify a core structure then we 
have examined the structure visually.

\begin{figure}[!htbp] \centering
  \begin{tabular}{@{}ccc@{}}
    \includegraphics[width=2.4cm,clip=true]{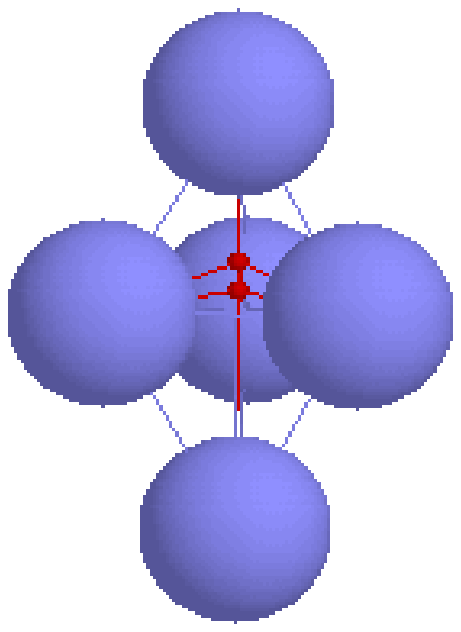} &
    \includegraphics[width=2.4cm,clip=true]{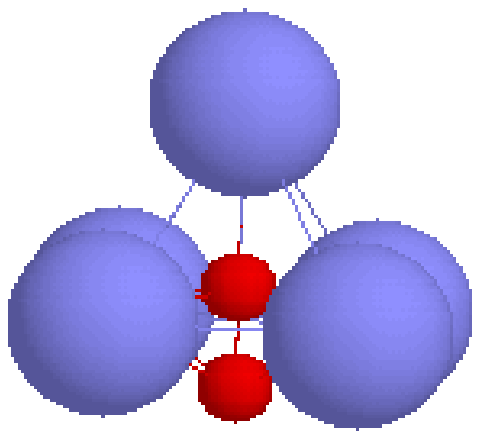} &
    \includegraphics[width=2.4cm,clip=true]{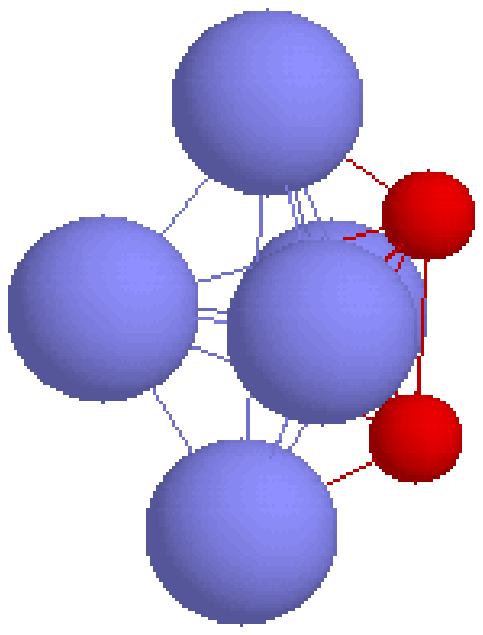} \\
    (5.1) & (5.2) & (5.3) \\
    \includegraphics[width=2.8cm,clip=true]{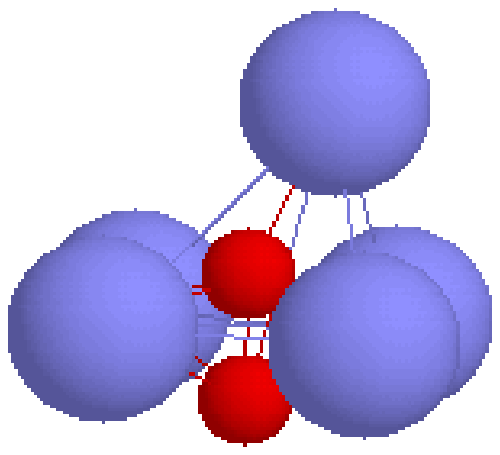} & 
    \includegraphics[width=3.0cm,clip=true]{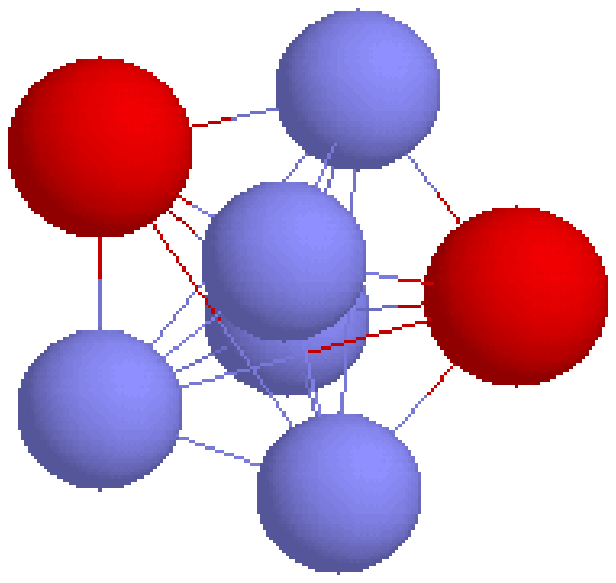} & \\
    (5.4) & (5.5) & \\
    \includegraphics[width=3.2cm,clip=true]{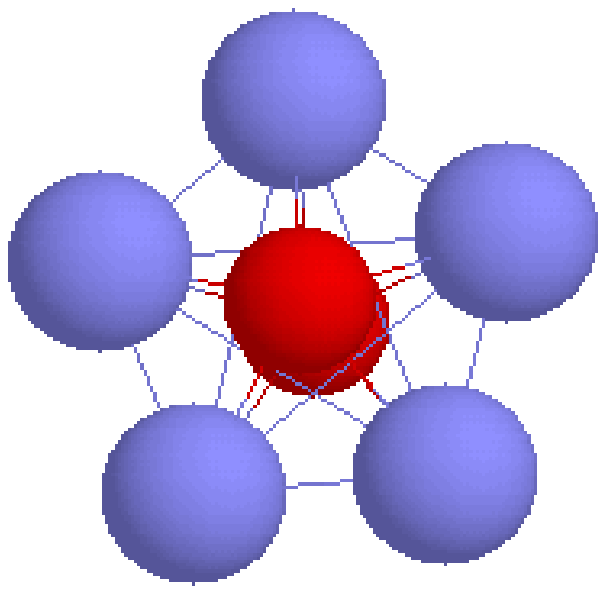} & 
    \includegraphics[width=3.0cm,clip=true]{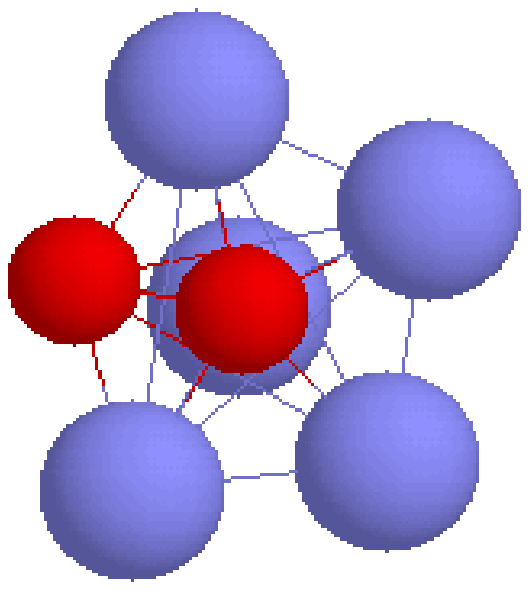} & \\
    (5.6) & (5.7) & \\
        \end{tabular}
\caption{\label{fig:coreE52}
Plots of $\mathrm{X_5Y_2}$ structures for selected
$\mathrm(\sigma,\epsilon)$ values. The decimal number for each figure denotes
the corresponding $\mathrm(\sigma,\epsilon)$ domain in 
Fig.~\ref{fig:corePES52}.} 
\end{figure}

Selected cluster structures
illustrating the core arrangements corresponding to various 
($\sigma$,$\epsilon$) values are shown in Fig.~\ref{fig:coreE52}. We see from
Figs.~\ref{fig:corePES52} and ~\ref{fig:coreE52} that the $\mathrm{X_5Y_2}$
cluster exhibits core X-atom structures that include trigonal bipyramidal,
planar and square pyramidal core geometries. Of these, only the trigonal
bipyramidal form is stable in the parent $\mathrm X_5$ system. This
illustrates that a suitable choice of the ($\sigma$,$\epsilon$) parameters
can controllably induce core geometries not present as stable minima in
the bare cluster. For example, the square pyramid core structure, seen
in Fig.~\ref{fig:coreE52}.2 as a stable system, corresponds to a transition
state in the bare $\mathrm X_5$ cluster. 

\begin{figure*}
  \begin{tabular}{@{}cc@{}}
    \includegraphics[width=8.5cm,clip=true]{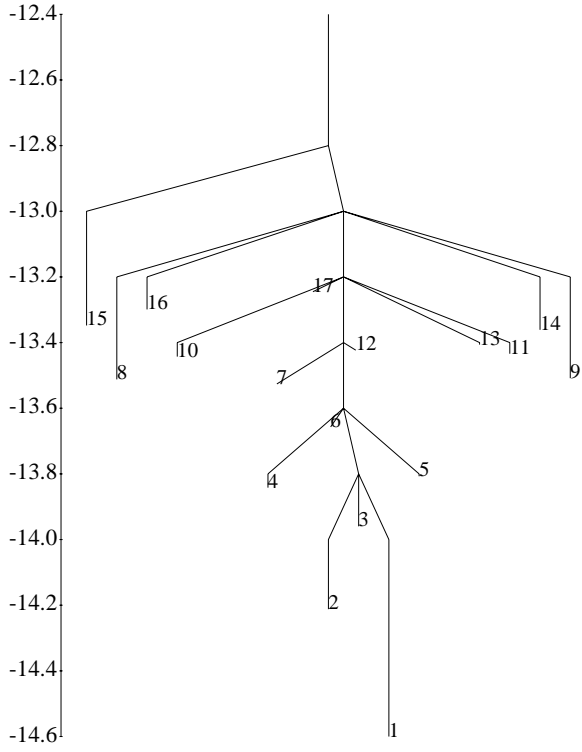} &
    \includegraphics[width=8.5cm,clip=true]{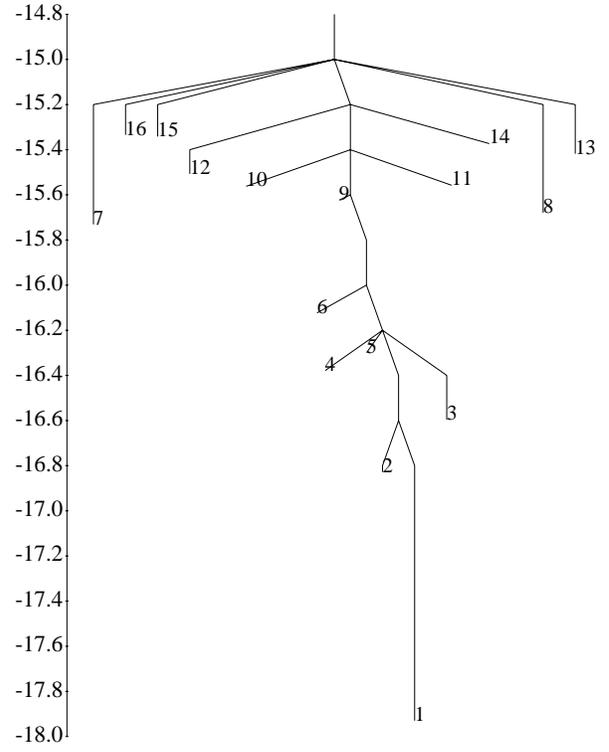} \\
    (a) & (b) \\
    \includegraphics[width=8.5cm,clip=true]{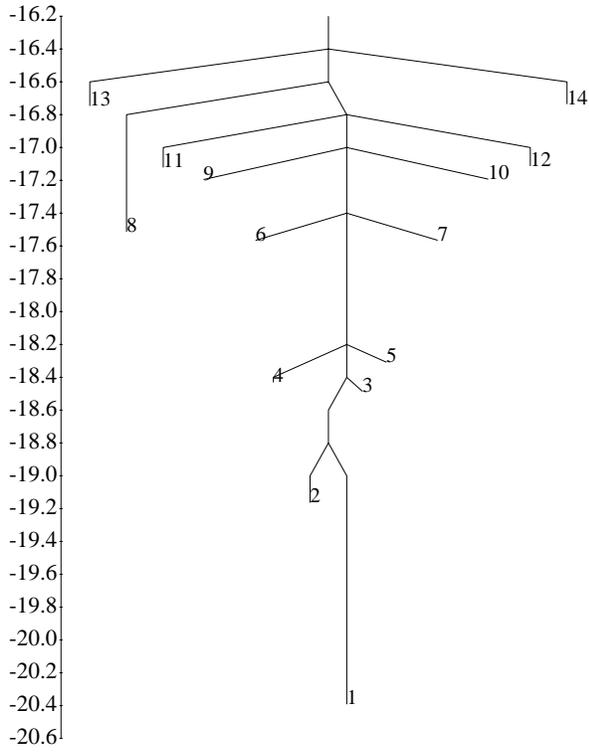} &
    \includegraphics[width=8.5cm,clip=true]{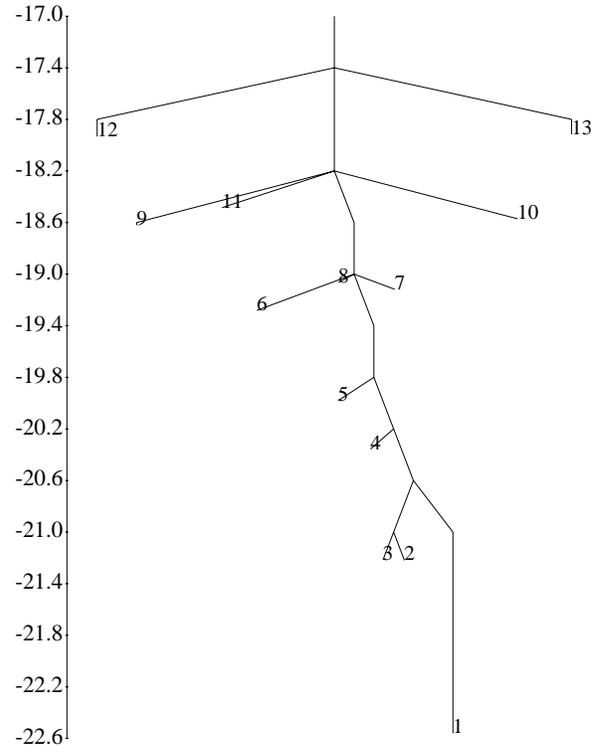} \\
    (c) & (d) \\
        \end{tabular}
\caption{\label{fig:DisCon52}
Disconnectivity graph for $\mathrm{X_5Y_2}$
$\mathrm(\sigma,\epsilon)$ values demonstrating that we can control
barriers for the selected inherent structures. The energy scale is in
units of $\epsilon_{_{XX}}$. The $\mathrm(\sigma,\epsilon)$ values for
panels (a--d) are (0.4,0.5), (0.4,1.0), (0.4,1.5) and (0.4,2.0), 
respectively.}
\end{figure*}

Fig.~\ref{fig:DisCon52} represents the $\mathrm{X_5Y_2}$ cluster at
four points in Fig.~\ref{fig:corePES52} defined by the ($\sigma$,$\epsilon$)
coordinates (0.4,0.5), (0.4,1.0), (0.4,1.5) and (0.4,2.0). 
Here the pairs of coordinates correspond 
to (a), (b), (c) and (d) part of Fig.~\ref{fig:DisCon52}, respectively.
In other words, we keep value of $\sigma$=0.4 fixed, while increasing
the value of $\epsilon$.
Each disconnectivity graph shows all inherent structures available to
the system for the given ($\sigma$,$\epsilon$) values. The global minimum
of each system is labeled by number 1 and contains as a recognizable
component the square pyramid core structure (see Fig.~\ref{fig:coreE52}.2).
In Fig.~\ref{fig:DisCon52}.a the square pyramid core structure is
connected to two inherent structures, labeled by 2 and 3, by pathways 
whose energies do not exceed $-13.8$ (in units of $\epsilon_{_{XX}}$).
Since isomer \# 2 contains the same core structure 
(the square pyramid) as the global minimum the corresponding isomerization
thus does not lead to a change in the core structure of the cluster. For
present purposes, therefore, the barrier that connects them is not a 
``relevant'' barrier. The relevant barriers are those that
connect inherent structures that contain different core structures.
The inherent structure \# 3 contains as the core structure a (distorted) 
trigonal bipyramid (see Fig.~\ref{fig:coreE52}.1). 
Therefore, the isomerization barrier that connects the
inherent structure \# 3 with global minimum is the lowest relevant 
isomerization barrier and
its value is $\Delta$E$_{1,3}$=0.986$\epsilon_{_{XX}}$.
Fig.~\ref{fig:DisCon52}.b and Fig.~\ref{fig:DisCon52}.c show that
increasing the value of $\epsilon$ increases isomerization barriers
that connect inherent structure \#1 (the square pyramid core structure)
with inherent structure \#2 (the distorted trigonal bipyramid core structure).
Numerically, these barriers are 
$\Delta$E$_{1,2}$=1.227$\epsilon_{_{XX}}$ and
$\Delta$E$_{1,2}$=1.431$\epsilon_{_{XX}}$, respectively.
In Fig.~\ref{fig:DisCon52}.d the square pyramid core structure is
connected by an isomerization barrier of 
$\Delta$E$_{1,2}$=1.647$\epsilon_{_{XX}}$
with two (almost degenerate in energy) distorted trigonal bipyramid
core structures.
As illustrated in Fig.~\ref{fig:DisCon52} and discussed above,
the barriers that determine the isomerization
kinetics of these newly induced structures are sensitive to the
($\sigma$,$\epsilon$) values and can thus be at least partially controlled.
These two simple results are specific demonstrations of goals 1 and 3
stated above.


\begin{figure}
\includegraphics[clip=true,width=8.5cm]{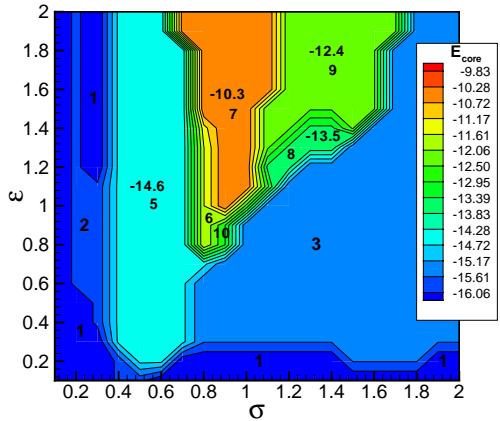}
\caption{\label{fig:corePES73}
$\mathrm E_{core}(\sigma,\epsilon)$ for $\mathrm{X_7Y_3}$.
Format for the plot is
the same as in Fig.~\ref{fig:corePES52}.}
\end{figure}

\begin{figure}[!htbp] \centering
  \begin{tabular}{@{}cc@{}}
    \includegraphics[width=3.4cm,clip=true]{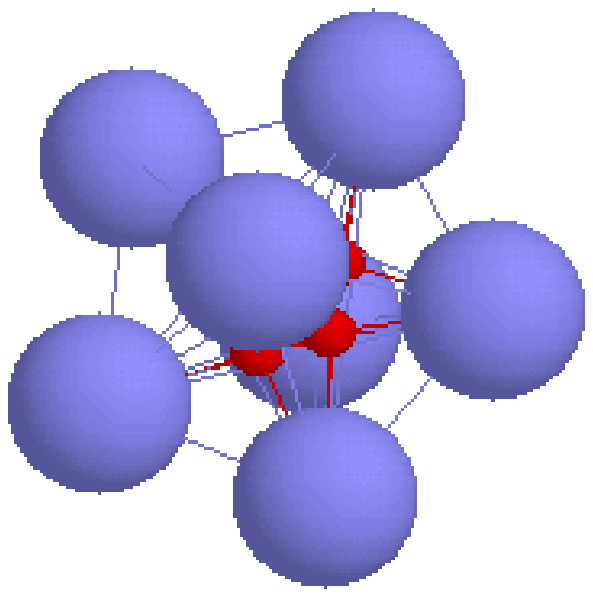} &
    \includegraphics[width=3.4cm,clip=true]{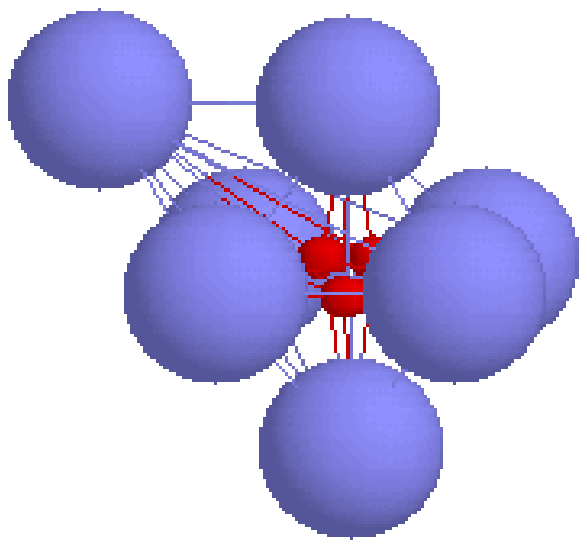} \\
    (8.1) & (8.2) \\
    \includegraphics[width=3.4cm,clip=true]{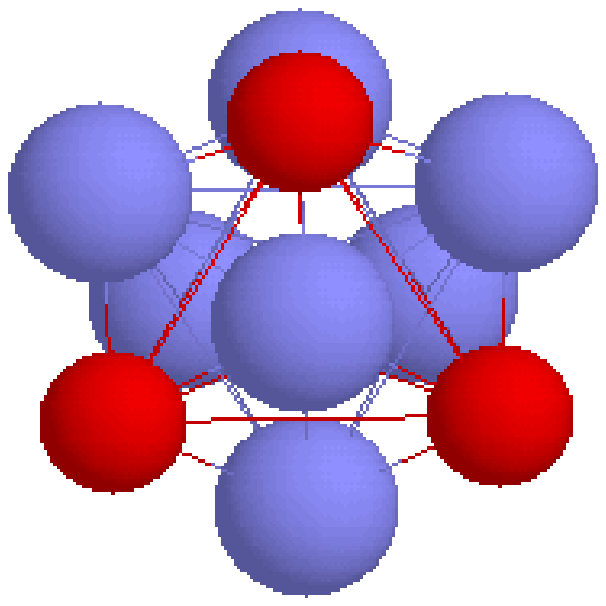} &
    \includegraphics[width=2.8cm,clip=true]{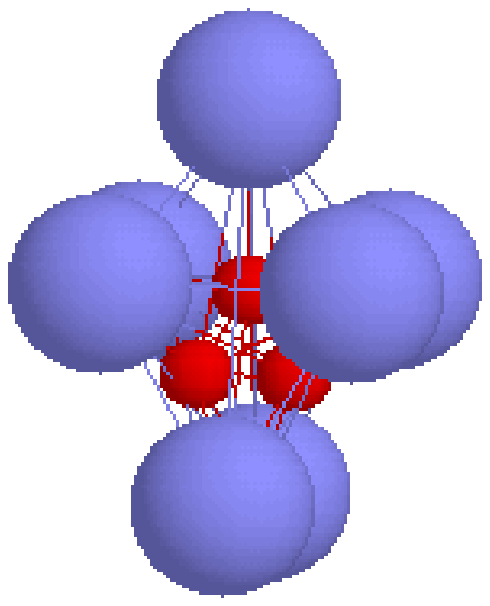} \\
    (8.3) & (8.5) \\
    \includegraphics[width=3.4cm,clip=true]{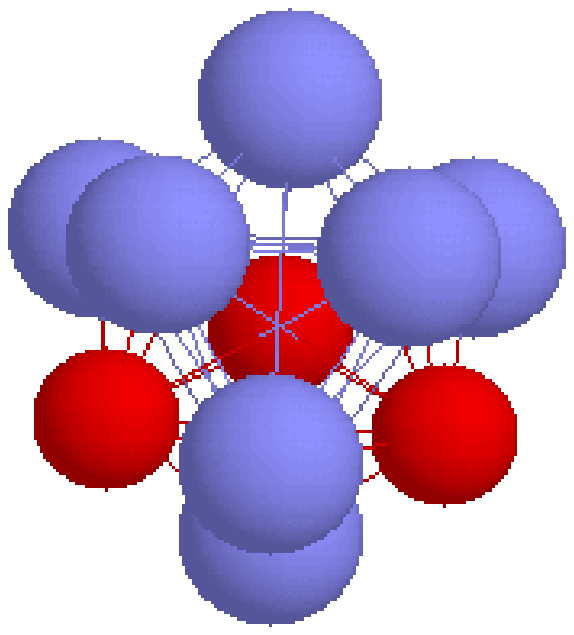} &
    \includegraphics[width=3.6cm,clip=true]{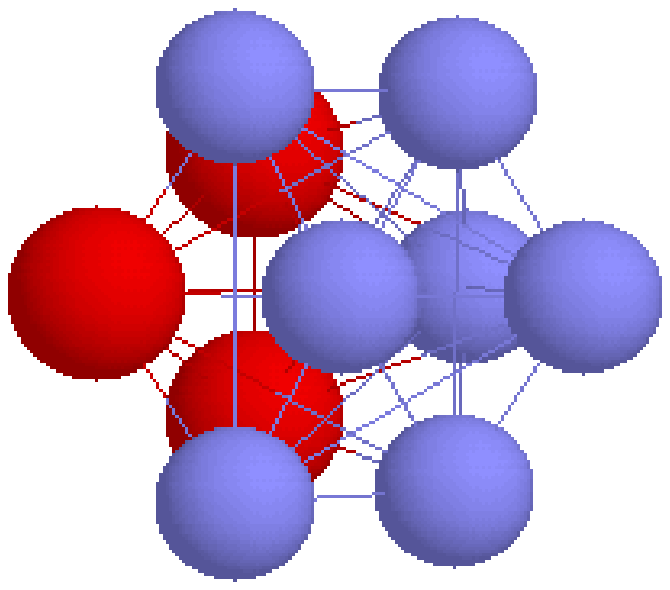} \\
    (8.6) & (8.8) \\
        \end{tabular}
\caption{\label{fig:coreE73}
Plots of selected $\mathrm{X_7Y_3}$ structures for various
$\mathrm(\sigma,\epsilon)$ values identified in Fig.~\ref{fig:corePES73}.
The number of the structures correspond to the regions labeled in
Fig.~\ref{fig:corePES73}. Note that many of the core structures for
these systems are not stable energy structures of the bare
$\mathrm X_7$ system.}
\end{figure}

\begin{figure*}
  \begin{tabular}{@{}cc@{}}
    \includegraphics[width=8.5cm,clip=true]{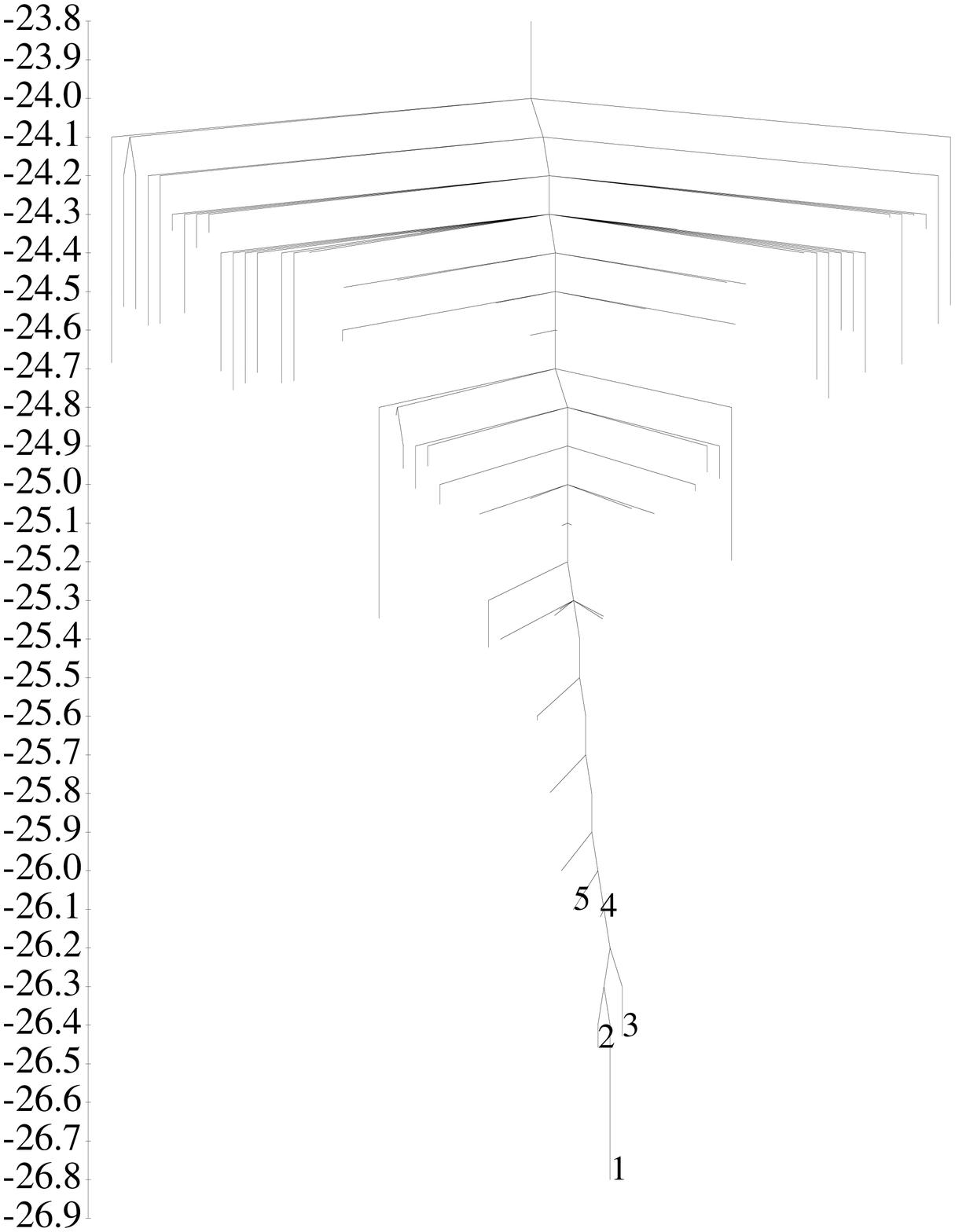} &
    \includegraphics[width=8.5cm,clip=true]{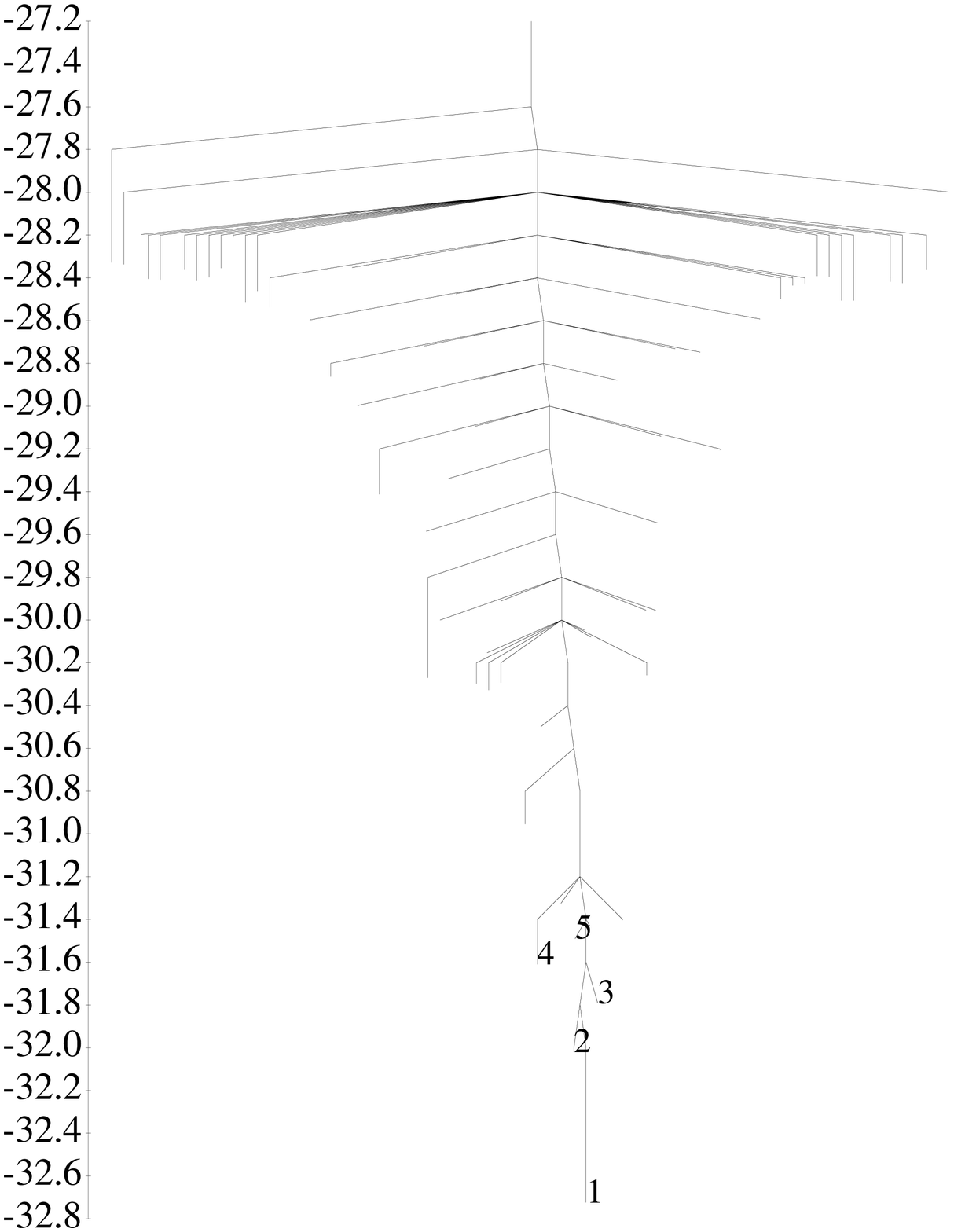} \\
    (a) & (b) \\
    \includegraphics[width=8.5cm,clip=true]{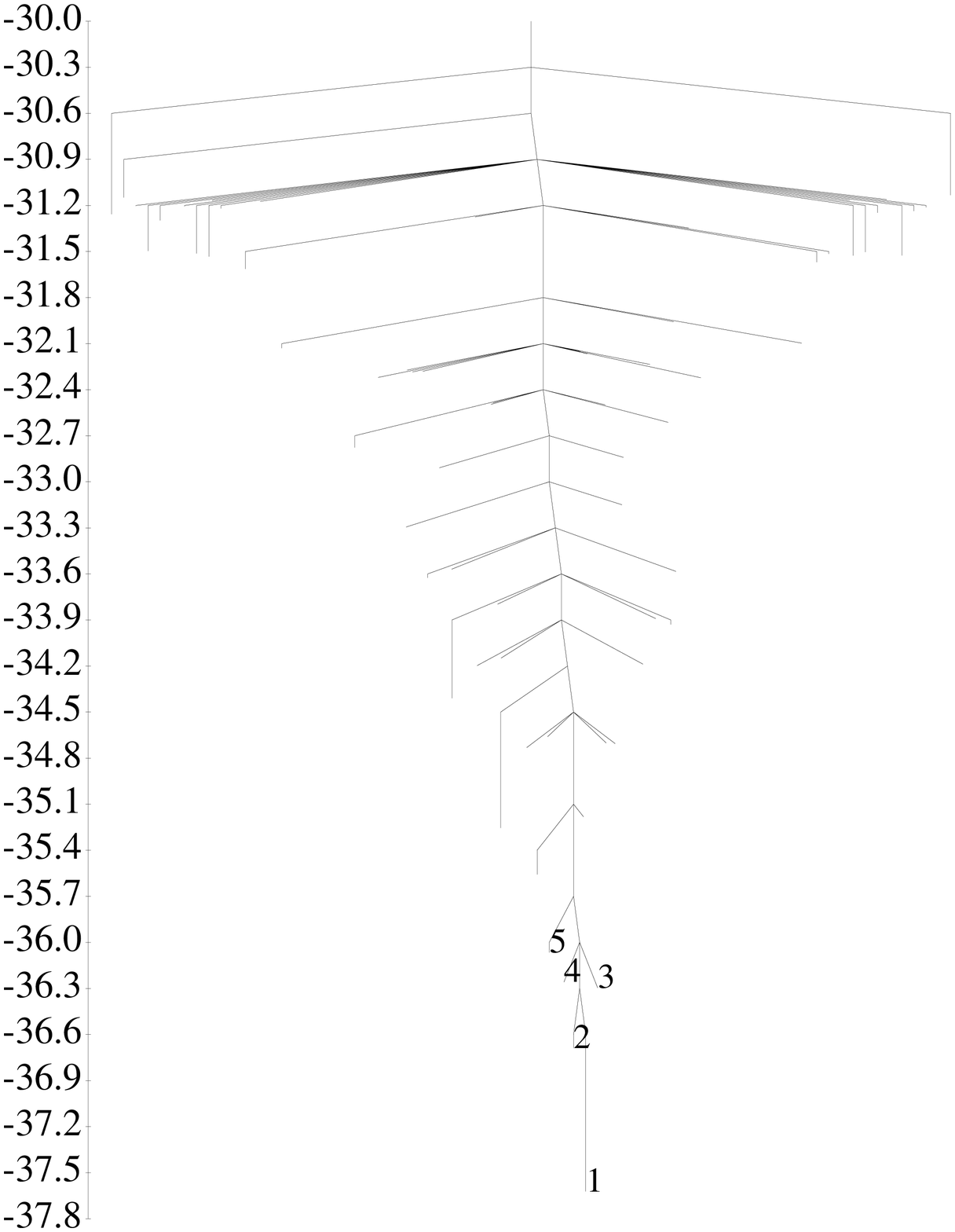} &
    \includegraphics[width=8.5cm,clip=true]{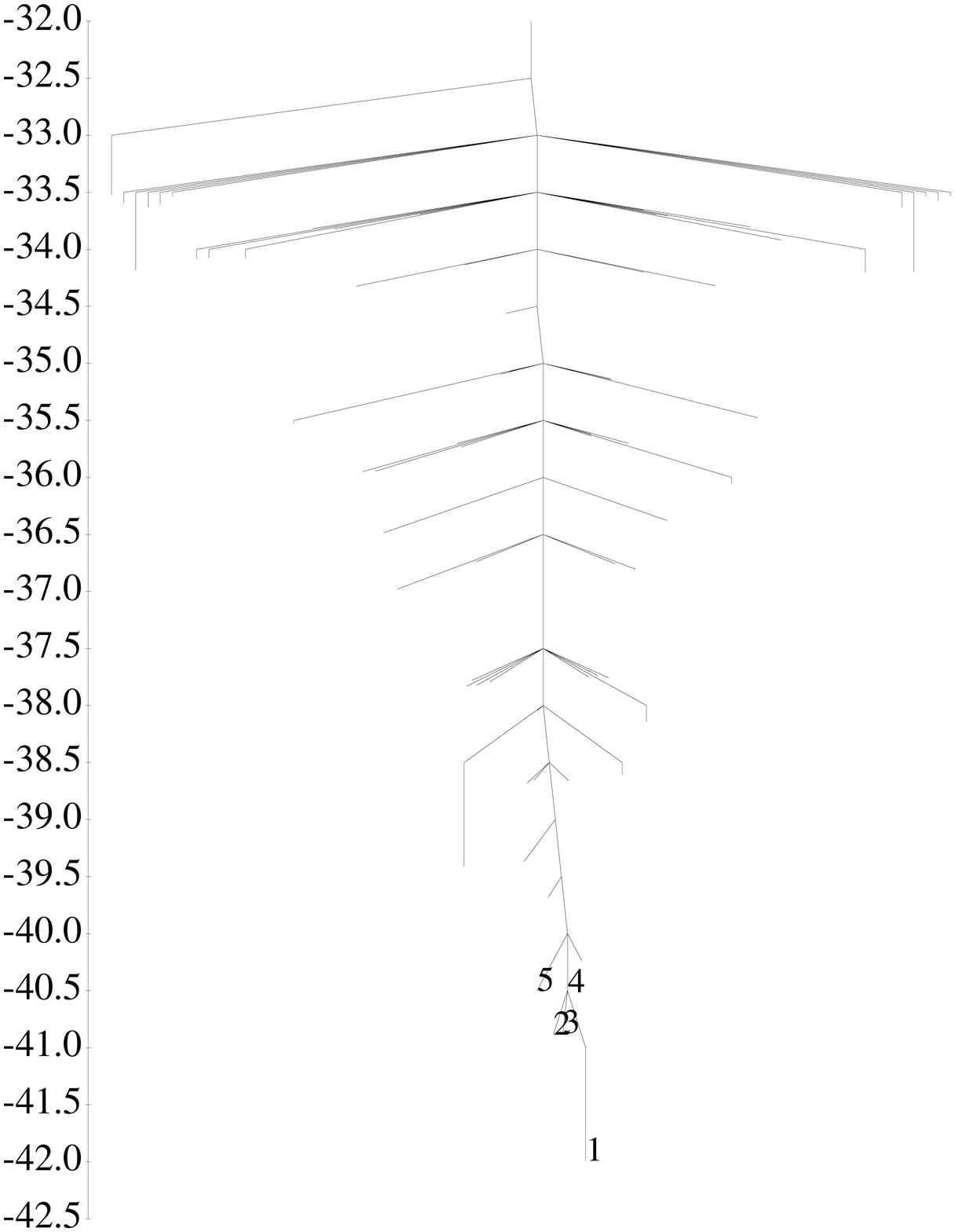} \\
    (c) & (d) \\
        \end{tabular}
\caption{\label{fig:DisCon73}
Disconnectivity graph for $\mathrm{X_7Y_3(\sigma,\epsilon)}$
values demonstrating that we can control barriers for the selected
inherent structures. The energy scale is in units of $\epsilon_{_{XX}}$.
Only branches leading to the 70 lowest-energy minima are shown.}
\end{figure*}

As a second illustration, we consider mixed clusters of the type
$\mathrm{X_7Y_3}$. This  system builds upon a parent, seven-atom, ``magic
number'' system known to exhibit a set of four, energetically distinct
inherent structures. The core inherent structures and associated energies
for the stable $\mathrm X_7$ inherent structures are presented in
Fig.~\ref{fig:core7}. Figure~\ref{fig:corePES73}, a 
$\mathrm(\sigma,\epsilon)$ contour plot of the core-atom potential 
energies of the lowest total energy $\mathrm{X_7Y_3}$ clusters, again
reveals the presence of definite ``core-phases''.
As illustrated in
Fig.~\ref{fig:coreE73}, some of these regions correspond to various
core structures present in the parent $\mathrm X_7$ system while others
correspond to new structures not seen in the original, single-component
cluster. We can see from Figs.~\ref{fig:corePES73}$-$\ref{fig:DisCon73}
that the impurity Y-atoms provide us with significant control over the 
relative ordering of the core energies of the parent $\mathrm X_7$
system. Specifically, by choosing an appropriate range of
$\mathrm(\sigma,\epsilon)$ values, we can generate $\mathrm{X_7Y_3}$
clusters in which the lowest (total) energy inherent structure can
have core structures that are either pentagonal bipyramid, capped octahedral,
or bicapped trigonal bipyramidal in nature. Moreover, since we can
manipulate the isomerization barriers in these systems, we can at least
partially stabilize clusters that exhibit selected core structures with
respect to isomerization. This is illustrated in Fig.~\ref{fig:DisCon73}.

Figures~\ref{fig:DisCon73}.a~--~\ref{fig:DisCon73}.d represent the 
$\mathrm{X_7Y_3}$ cluster at four points in Fig.~\ref{fig:corePES73} with
$\mathrm{X_7Y_3(\sigma,\epsilon)}$ coordinates (0.4,0.5), (0.4,1.0), 
(0.4,1.5) and (0.4,2.0), respectively.
The number of inherent structures available to the $\mathrm{X_7Y_3}$ 
cluster varies from more than 800 in Fig.~\ref{fig:DisCon73}.a to 400
in Fig.~\ref{fig:DisCon73}.d. Since we are primarily interested in
energetically low-lying inherent structures we show only lowest 70
inherent structures. The global minimum of each system is labeled by
number 1 and contains as a recognizable component the core structure
shown in Fig.~\ref{fig:coreE73}.5. We should mention that for a given
range of $\sigma$ and $\epsilon$ values, ($\sigma, \epsilon$)$\in$ [0.1,2.0],
we have not been able to find a global minimum that would contain as
a recognizable component inherent structure \# 4 of the parent $\mathrm{X_7}$
cluster (see Fig.~\ref{fig:core7}.d). This is the reason why none of the
domains in Fig.~\ref{fig:corePES73} is labeled by number 4. 
In Fig.~\ref{fig:DisCon73}.a the global minimum, the core structure \# 5
(see Fig.~\ref{fig:coreE73}.5), is linked to inherent structure \# 2
which contains the (distorted) capped octahedron core structure (see
Fig.~\ref{fig:coreE73}.2).
The isomerization barrier between them is 
$\Delta$E$_{1,2}$=0.494$\epsilon_{_{XX}}$.
Fig.~\ref{fig:DisCon73}.b and Fig.~\ref{fig:DisCon73}.c show that
increasing the value of $\epsilon$ increases isomerization barriers,
that connect inherent structure \#1 (see core structure in
Fig~\ref{fig:coreE73}.5) 
with inherent structure \#3 and \#2
(a distorted capped octahedron core structure), respectively.
Numerically, these barriers are
$\Delta$E$_{1,3}$=0.975$\epsilon_{_{XX}}$ and
$\Delta$E$_{1,2}$=1.136$\epsilon_{_{XX}}$, respectively.
The inherent structure \#2 in Fig.~\ref{fig:DisCon73}.b contains the same
core structure as the global minimum and, therefore, has not been considered
relevant for the isomerization (see explanation above for
Fig.~\ref{fig:DisCon52}.a).
In Fig.~\ref{fig:DisCon52}.d the core structure \# 5 is
connected to two, energetically almost degenerate, inherent structures
labeled by 2 and 3, by pathways whose energies do not exceed $-40.5$ (in
units of $\epsilon_{_{XX}}$). Similarly to the case of 
Fig.~\ref{fig:DisCon52}.a, the isomer \# 2 contains the same core structure
as the global minimum so the corresponding isomerization does not lead
to a change in the core structure of the cluster. The barrier that links
them is not a relevant barrier. The inherent structure \# 3 contains
as the core structure a distorted capped octahedron. Therefore, the
isomerization barrier which connects the inherent structure \# 3 with
global minimum is the lowest relevant isomerization barrier and its value
is $\Delta$E$_{1,3}$=1.304$\epsilon_{_{XX}}$.

\section{Conclusions} \label{sec:conclude}

In the present work, we have considered the general task of
altering core cluster structures.  We are, in effect, attempting to turn the
logic of the minimization problem upside down.  Rather than seeking the
global minimum of complex potential energy surfaces, we are instead attempting
to exploit what has been learned about the general minimization problem to
controllably alter core cluster structures.  Specifically, we are examining
the extent to which we can induce new core geometries as well as reorder and
stabilize existing, higher-lying, local core structures.

Our approach, in the present discussion, has been thermodynamic in nature.  We
have utilized selected adatoms to effect our desired core cluster
modifications.  We have presented results for two simple binary cluster
examples, the $\mathrm{X_5Y_2}$ and $\mathrm{X_7Y_3}$ systems, to validate 
our approach.

We speculate that there are at least two important directions for future
theoretical development of the present ideas.  One direction will be to
explore the use of more complex adsorbates to achieve selected core cluster
structures.  One could, for example, imagine using ``exterior'' methods in 
which
encapsulating agents of well-defined geometries were utilized to induce desired
core structures.  Alternatively, ``interior'' approaches in which complex
objects, perhaps even previously engineered clusters, could be utilized as 
``seeds'' or ``templates'' to produce a desired structure in the 
surrounding cluster
(either globally or locally).  Another important direction will be to explore
the extent to which previously engineered cluster structures can be assembled
using ``cluster assembled materials'' methods to produce larger scale,
macroscopic structures.  If this proves possible, it would seem to offer an
important direction in the production of novel materials starting from
synthetic precursors whose core structures and properties are highly varied 
and are under user control.

{\bf Acknowledgment}

The authors acknowledge support from the National Science Foundation
through awards No. CHE-0095053. and CHE-0131114. They would also like
to thank Dr. M. Miller for helpful discussions and for his gracious
assistance with respect to the preparation of the disconnectivity graphs
in the present paper.

\bibliography{sabo}


\end{document}